\documentstyle[preprint,prl,aps]{revtex}

\begin{document}
\preprint{\vbox{\hbox{April 1994}\hbox{IFP-701-UNC}\hbox{VAND-TH-94-6}}}
\draft
\title{Minimal Family Unification}
\author{\bf P. H. Frampton$^{(a)}$ and T. W. Kephart$^{(a,b)}$}
\address{(a)Institute of Field Physics, Department of Physics and Astronomy,\\
University of North Carolina, Chapel Hill, NC  27599-3255}
\address{(b)Department of Physics and Astronomy,Vanderbilt University,
Nashville, TN 37235\footnote{Permanent address}}
\maketitle

\begin{abstract}
It is proposed that there exist, within a new $SU(2)^{'}$, a gauged discrete
group $Q_6$ (the order 12 double dihedral group) acting as a family symmetry.
This nonabelian finite group can explain
hierarchical features of families, using an assignment for quarks and leptons
dictated by the requirements of anomaly cancellation and of no additional
quarks.
\end{abstract}
\newpage
Finite groups play a major role in physics. To cite just a few
of very many examples, they are central to molecular orbitals
just as the crystallographic groups are in solid-state physics;
the discrete symmetries C, P and T and their violation have made a
profound impact on our understanding of quantum field theory.

The masses and mixings of the quark-lepton families make up a majority
of the parameters in the existing framework of particle theory and it is
natural to systematize the observed hierarchies between such parameters
by using a Family Symmetry (FS) under which the families transform in a
nontrivial fashion.
Such an FS might be a continuous Lie group or, more economically, a
finite group. A finite group FS may be conveniently constructed as a
subgroup of an anomaly-free gauged Lie group.

Amongst finite groups\cite{books}, the non-abelian examples have the advantage
of non-singlet irreducible representations which can be used to inter-relate
families. Which such group to select is based on simplicity: the minimum
order and most economical use of representations\cite{guts}. The smallest
non-abelian finite group is $S_3$ ($\equiv D_3$), the symmetry of an
equilateral triangle. This group initiates two infinite series, the $S_N$
and the $D_N$. Both have elementary geometrical significance since the
symmetric permutation group $S_N$ is the symmetry of the N-plex in N
dimensions while $D_N$ is the symmetry of the planar N-agon in 3 dimensions.
As a family symmetry, the $S_N$ series becomes uninteresting rapidly
as the order and the dimensions of the representions increase. Only $S_3$
and $S_4$ are of any interest as symmetries associated with the particle
spectrum\cite{Pak}, also the order (number of elements) of the $S_N$ groups
grow factorially with N. The order of the dihedral groups increase only
linearly with N and their irreducible representations are all one- and
two- dimensional. This is reminiscent of the representations of the
electroweak $SU(2)_L$ used in Nature.

In the observed masses and mixings of quarks, the third family (especially
the top quark mass) is the most different. The FS must, as a first
requirement, single out this feature with the hope that details of the
first and second families will be amenable to study on the basis of
the FS framework.

Consider, to set the scene, using $D_7$ which has two singlet ($1$
and $1^{'}$) and three doublet $2_{(j)}$ ($1\leq j\leq 3$)
representations. The multiplication rules are:
\begin{equation}
1^{'}\times 1^{'} = 1 ; ~~~1^{'}\times 2_{(j)} = 2_{(j)}
\end{equation}
\begin{equation}
2_{(i)}\times 2_{(j)} = \delta_{ij} (1 + 1^{'}) + 2_{(min[i+j,N-i-j])}
+ (1 - \delta_{ij}) 2_{(|i - j|)}
\end{equation}
\noindent
with N = 7 (the above is valid for any odd N). This $D_7$ commutes with the
standard model group. It is natural to try an assignment such as:

$$\begin{array}{ccccccc}

\left( \begin{array}{c} t \\ b \end{array} \right)_{L} &
1 & \begin{array}{c} t_{R}~~~ 1 \\ b_{R} \hspace{0.2in}1^{'} \end{array} &
\left( \begin{array}{c} \nu_{\tau} \\ \tau \end{array} \right)_{L} & 1 &
\tau_{R} & 1^{'} \\

\left. \begin{array}{c} \left( \begin{array}{c} c \\ s \end{array} \right)_{L}
\\
\left( \begin{array}{c} u \\ d \end{array} \right)_{L} \end{array}  \right\} &
2_{(1)}
 &  \begin{array}{c} \left. \begin{array}{c} c_{R} \\ u_{R} \end{array}
\right\} 2_{(2)} \\
\left. \begin{array}{c} s_{R} \\ d_{R} \end{array} \right\} 2_{(2)} \end{array}
&
\left. \begin{array}{c} \left( \begin{array}{c} \nu_{\mu} \\ \mu \end{array}
\right)_{L} \\
\left( \begin{array}{c} \nu_{e} \\ e \end{array} \right)_{L} \end{array}
\right\} & 2_{(1)}
& \left. \begin{array}{c} \mu_{R} \\ \\ e_{R} \end{array} \right\} & 2_{(2)}

\end{array}$$

Giving vacuum values ( VEVs ) to the Higgs doublets in the $1'$
and $2_{(3)}$ of $D_7$ then gives mass matrices with hierarchical
treatment of the third family. The $t$ acquires a mass without breaking
$D_7$\cite{rootN} (Any $D_N$ model with less than two {\bf 2}'s allows
unwanted mass and/or mixing terms, thus we must have $N \geq 7$). Let
us pause here to place the $D_N$ family symmetry
in a proper modern context. It is now known that any global symmetry
is violated by quantum gravity effects\cite{global}.
To avoid this problem it is necessary to gauge our
discrete symmetry. The simplest approach is to embed
$D_N$ in $O(3)$ and then gauge the $O(3)$. When we make this
choice it is necessary to have a mechanism of breaking
$SO(3)$ to $D_N$.  This can be easily arranged by the following
Higgs potential for N triplets of $SO(3)$.
\begin{equation}
V = \sum_{i=1}^{N} \sum_{p=1}^{N} (\phi_{i}.
\phi_{i+p} - v^{2} cos (2\pi p/N))^{2}
\end{equation}

But now, for $D_N$ to be properly gauged the particle spectrum must
fall into complete representations of $SO(3)$, otherwise
the theory may have chiral
anomalies\cite{ross}. As is easily seen the $D_N$ model above is flawed
for this reason and it is not difficult to show that the simultaneous
constraints of mass hierarchy and anomaly cancellation cannot be satisfied
for $D_7$. ( We have shown this statement
to be true for any $D_N$ FS model. ) The difficulty can be traced
to the fact that
there are only integer "angular momentum" representations in $SO(3)$, and
it is only possible then to use one {\bf 2} of $D_7$ unless more states
are added to the theory. Although adding many more states is a possibility
worth exploring we prefer here to stay as close as possible to a three family
standard model and search further for a discrete FS satisying more constraints.

To this end we consider the double dihedral groups $Q_{2N}$ (also called
the dicyclic groups\cite{books} ), of order 4N, which are the spinor
generalization of $D_N$ where now $Q_{2N}$ is embedded in $SU(2)$, the covering
group
of $SO(3)$. Specifically consider $Q_{6}$ where its representations are
$1$, $1^{'}$, $1^{''}$, $1^{'''}$, $2$ and $2_{S}$ with multiplication
table:
$$\begin{tabular}{||c||c|c|c|c|c|c||}   \hline
$1$ &$ 1$ &$ 1'$ & $1''  $ & $1'''$ &$ 2$ &$ 2_{S}$\\    \hline\hline
$1$  &$ 1$ &$ 1'$ &$ 1''$   &$ 1'''$ & $2$ &$ 2_{S}$ \\  \hline
$1'$ & & $1$ & $1'''$ & $1''$ &$ 2$ &$ 2_{S}$ \\    \hline
$1''$& & & $1'$ & $1$ & $2_{S}$ & $2$ \\   \hline
$1'''$& & & &$ 1'$ & $2_{S}$ &$ 2$ \\  \hline
$2$& & & & & $2 + 1 + 1'$ &$ 2_{S} + 1'' + 1'''$\\  \hline
$2_{S}$ & & & & & &$ 2 + 1 + 1'$   \\   \hline
\end{tabular}$$

There are $3^{5}$ ways of assigning the 5 triples of fermions with
common quantum numbers: $(u^{i},d^{i})_{L}$, $\overline{u}^{i}_{L}$,
$\overline{d}^{i}_{L}$, $(\nu^{i},l^{i})_{L}$, and $\overline{l}^{i}_{L}$
to one of the three anomaly-free sets
$1 +1 +1$, $1^{'} + 2$, $1 + 2_{S}$; note that $1^{'} + 2$ is the {\bf 3}
of $SU(2)^{'}$ and that $2_{S}$ is the {\bf 2}.
Of these many possibilities, interesting mass matrices and mixing angles
arise only from the assignment:

$$\begin{array}{ccccccc}

\left( \begin{array}{c} t \\ b \end{array} \right)_{L} &
1 & \begin{array}{c} t_{R}~~~ 1 \\ b_{R} \hspace{0.2in}1^{'} \end{array} &
\left( \begin{array}{c} \nu_{\tau} \\ \tau \end{array} \right)_{L} & 1 &
\tau_{R} & 1^{'} \\

\left. \begin{array}{c} \left( \begin{array}{c} c \\ s \end{array} \right)_{L}
\\
\left( \begin{array}{c} u \\ d \end{array} \right)_{L} \end{array}  \right\} &
2_{S}
 &  \begin{array}{c}  \begin{array}{c} c_{R}~~~1 \\ u_{R}~~~1 \end{array}  \\
\left. \begin{array}{c} s_{R} \\ d_{R} \end{array} \right\} 2 \end{array} &
\left. \begin{array}{c} \left( \begin{array}{c} \nu_{\mu} \\ \mu \end{array}
\right)_{L} \\
\left( \begin{array}{c} \nu_{e} \\ e \end{array} \right)_{L} \end{array}
\right\} & 2_{S}
& \left. \begin{array}{c} \mu_{R} \\ \\ e_{R} \end{array} \right\} & 2

\end{array}$$

These assignments are free from the global $SU(2)^{'}$ anomaly since there
is a total even number ( eight ) of the $2_{S}$ = {\bf 2} of $SU(2)^{'}$.
For the $(SU(2)^{'})^2Y$ anomaly, taking the normalization $Q = T_3 + Y$,
and putting the quadratic Casimir for the ${\bf 2_S}$ as +1 and hence that for
the ${\bf 3}$ as +4,
the result is +8 from the quarks and leptons of the standard model.
It can be shown that without extending the particle spectrum
 no assignments under $Q_6$ cancels this final anomaly.  It is most
economically cancelled by adding leptons with contribution -8; for
example, we may add leptons which are vector-like under the standard model but
transform, for the
left-handed doublets, as two ${\bf 3}s$ and, for the right-handed doublets,
as $Q_6$ singlets. Such new leptons, between about $50$ GeV and $200$ GeV,
are the smoking gun for the $Q_6$ model. The quark sector remains unchanged.

Before continuing with the analysis of this model it is important to
note that we have not chosen the  $Q_6$ group at random, but have made a
systematic study of all finite groups $F$ of order $\leq 31$. There are 93
groups on this list, 45 of which are non-Abelian.  On analysis of the
representation content, product rules, and embeddings into
continuous groups, along with the phenomenological constraints that (i)
the top quark is an $F$ singlet and lighter fermions acquire masses in
sequential breaking of $F$ (e.g., $b$ and $\tau$ masses appear at the first
stage of $F$ breaking). (ii) No additional quarks are permitted
in the theory, and (iii) total anomaly freedom.  These rules are sufficient
to eliminate all but the groups $Q_{2N}$ and $T_d$ (of order 24), the spinor
version of the tetrahedral group.  All these embed in $SU(2)$ and of
them $Q_6$ is the minimal choice. (The details of the above analysis
are rather lengthy and will appear elsewhere.)

Continuing now with the $Q_6$ model,
the mass matrices U, D and L have the potential textures:

$$U = \left( \begin{tabular}{c|c}
$<2_S>$ & $ <2_S> $ \\  \hline
$<1>  $ & $ <1>   $
\end{tabular} \right)$$

$$D = \left( \begin{tabular}{c|c}
$<1''+1'''+2_S>$ & $ <2_S> $ \\  \hline
$<2>  $ & $ <1'>   $
\end{tabular} \right)$$

$$L = \left( \begin{tabular}{c|c}
$<1''+1'''+2_S>$ & $ <2_S> $ \\  \hline
$<2>  $ & $ <1'>   $
\end{tabular} \right)$$

\noindent
in a notation where the upper-left block is $2\times 2$ for the first
two families and we have designated the VEVs which contribute to the
different entries.

The observed fermion mass heirarchy can now be arranged using roughly
equal Yukawa coupling constants but with a heirarchy of $SU(2)_L$
Higgs VEVs that sequentially break the gauge symmetry.  First an
$SU(2)_L \times Q_6$ VEV $<2,1>$ gives mass to the top.
(Note that we rotate the U matrix so that there is only
one diagonal entry for the top quark, and no mixing, by redefinition
of the fields; we follow similar procedures
throughout the chain of symmetry breaking.)

The gauged ancestral $SU(2)'$ is broken at a scale $(v_Q)$ to $Q_6$ using a
potential with a set of three triplet irreps of $SU(2)'$; this is
similar to Eq(3).  This breaking can also be accomplished with a third
rank symmetric tensor (a $\bf{7}$) of $SU(2)'$.
The scale  $v_Q$ is restricted from below by
the suppression of rare processes such as $K \rightarrow \pi \mu e$ with
branching ratio $10^{-10}$. This implies that $v_Q$ is at least $ 100 $
TeV, although the associated massive gauge bosons could be
lighter than this scale if ${g_2}'$ is sufficiently small. Without
further unification, ${g_2}'$ is a free parameter which could be so tiny
that these new particles are within reach of future colliders although
if ${g_2}'$ is comparable to ${g_2}$ they would be at O(100TeV) in mass.

The structure of the quark and lepton
mass matrices implies that we can leave $Q_6$ unbroken
down to a scale comparable to the $b$ and $\tau$ masses. At that scale $(v_b)$
$Q_6$ is broken by $< \bf{1}'>$ to $Z_6$.
Now an $SU(2)_L$ singlet VEV $<1,2_S>$ breaks $Z_6$ to $Z_2$
(which we note is not an entry in the texture matrices for U, D, and L above
since these entries are all of the form $<2$, irrep of $Q_6>$)
allows the charm quark to acquire mass at 1-loop with top in the loop.
A tree level mass for $s$ and $\mu$ comes via any of the three choices
$<2,1''>$, $<2,1'''>$, and $<2,2_S>$ (see below).  After unitary rotations
for the first two of these choices (both of which complete the breaking
of $Q_6$), $u$, $d$, and $e$
remain massless at tree level but  acquire light
masses at the next order in perturbation theory.  (Masses for these
particles can also be arranged through soft VEVs or a modest Yukawa
hierarchy.)

The initial stages of symmetry breaking is summarized by the diagram:
\begin{equation}
SU(2)_{L} \times U(1)_{Y} \times SU(2)'
\stackrel{v_Q(\bf{1,3})}{\longrightarrow}
 SU(2)_{L} \times
U(1)_{Y} \times Q_{6}
\stackrel{v_t(\bf{2,1})}{\longrightarrow} U(1)_{em} \times Q_{6}
\end{equation}
\begin{equation}
\stackrel{v_b(\bf{2,1'})}{\longrightarrow} U(1)_{em} \times Z_{6}
\end{equation}

At this stage only the third family of quarks and leptons gain their
(large) masses while the first two families remain massless. This is
the first step of the hierarchy.

Breaking of a discrete symmetry like $Q_6$ would generally lead to
unacceptable domain walls, but we may add a soft $Q_6$ breaking terms like
$m \phi \cdot (\phi \times \phi)$ to the potential, to avoid wall formation.
On the other hand, such walls would be acceptable if the distortion of the
cosmic background radiation due to the walls is sufficiently small.  We find
$\delta T/T \leq 10^{-4}$ if the
 self-coupling
$\lambda$ of the $\bf {1^{'}}$ field satisfies $\lambda \leq
10^{-5}$\cite{zel}.

The breaking of $Z_{6}$ occurs in two stages, the ordering of which
is a subtle problem which will determine the details of the masses
and mixings in the first and second families, and of their mixings
with the third family. The two possible chains of symmetry breaking
are:

$$\begin{array}{ccc}
$$ Z_{6}&\stackrel{v_c(\bf{2,2})}{\longrightarrow}& Z_{2}$$\\
$$   &\stackrel{v_s(\bf{2,1^{''} + 1^{'''}})}{\longrightarrow}& nothing.$$
\end{array}$$
and
$$\begin{array}{ccc}
$$ Z_{6} &\stackrel{v_s(\bf{2,1^{''} + 1^{'''}})}{\longrightarrow}& Z_{3}$$\\
 $$    & \stackrel{v_c(\bf{2,2+2_{S}})}{\longrightarrow}& nothing.$$
\end{array}$$

As stated above the
$c$ quark acquires mass from a radiative one-loop correction associated
with a VEV for a $({\bf 1,2_S})$ under $SU(2)_L \times SU(2)^{'}$, and this
VEV can also particapate in $Z_6$ breaking.

{}From the form of the mass
matrices, the VEV of $<2, 2_{S}>$ can give masses
to the second
family states $(s,\mu)$, and provide a mixings with the
third family. The VEVs of
$<2,1^{''} + 1^{'''}>$, expected to be somewhat smaller,
differentiate D, L from U. Although D and L have similar structures, it
does not necessarily imply any proportionality in these masses
(in D and L) because there are many independent (though roughly of
the same order of magnitude since we have enforced technical
naturalness) Yukawa coupling
coefficients. It will be interesting if the phases of the Yukawa couplings
themselves can be further constrained by the gauged $Q_6$ symmetry or
perhaps in a SUSY version of the theory.

To summarize, we have a minimal family unification based on the dicyclic group
$Q_6$ of order twelve\cite{groups}.  The model is minimal in that there
are no new quarks added to the theory\cite{SU3}.  The spectrum of the theory
is constrained to fall into complete representations of $SU(2)'$.
Since $SU(2)'$ has no chiral anomaly, the $Q_6$ theory is also chiral
anomaly free.  In addition we require an even number of $SU(2)^{'}$
fermion doublets to avoid a global $SU(2)'$
anomaly.
These requirements
constrains the $Q_6$ spectrum to the point that all models of
this type can be classified.  Finally, requiring that the top quark be allowed
to get a $Q_6$ invariant mass, that
the model is free of mixed anomalies,
and that the third family be unmixed with the
first two families at the $Z_6$ level completely fixes the model, and
predicts new leptons lighter than the top quark.

One of us (T.W.K.) thanks the members of the Institute of Field Physics
at UNC-Chapel Hill for their generous hospitality while this work was
in progress.  This work was supported in part by the U.S. Department of
Energy under Grants DE-FG05-85ER-40219 and DE-FG05-85ER-40226.


\begin{references}
\bibitem{guts}
Another (non-minimal by the above definition) route to discrete family
symmetry is through the previous $SU(N)$ family symmetry models\cite{SUN}
. Here one can consider the breaking pattern
\begin{equation}
SU(N)\rightarrow SU(5)\times SU(5-N)\times U(1)\rightarrow SU(5)\times G
\end{equation}
where G is a discrete group. Such models typically have more than just
three complete families until $G$ is broken. Although worthy of further
study, these models do not in general satisfy our minimality condition
on the fermion spectrum.
\bibitem{rootN}
Note that even though N 3-vector VEVs are needed to break $SO(3)$ to $D_N$
the top Yukawa has a factor $N^{-1/2}$ suppression as can be seen by
comparing the t and W masses. The b acquires its mass from breaking
to $Z_7$. The lighter quarks obtain mass from breaking of $Z_7$. The reversed
hierarchy for mixing angles is qualitatively explained since $\Theta_{12}$
receives a contribution from breaking $Z_7$ while the (smaller) angles
$\Theta_{23}$ and $\Theta_{13}$ do not.
\end{references}
\end{document}